\documentclass[pdflatex,sn-nature]{sn-jnl}
\usepackage{type1cm}
\usepackage{mathrsfs}
\usepackage{makecell}
\usepackage{svg}
\usepackage{booktabs} 
\usepackage{graphicx}%
\usepackage{multirow}%
\usepackage{amsmath,amssymb,amsfonts}%
\usepackage{amsthm}%
\usepackage[title]{appendix}%
\usepackage{xcolor}%
\usepackage{textcomp}%
\usepackage{manyfoot}%
\usepackage{booktabs}%
\usepackage{algorithm}%
\usepackage{algorithmicx}%
\usepackage{algpseudocode}%
\usepackage{listings}%
\usepackage{gensymb}
\usepackage[nolist]{acronym}
\usepackage{lineno}
\usepackage{fix-cm}

\begin{acronym}
\acro{CDI}{Coherent X-ray Diffraction Imaging}
\acro{CNN}{Convolutional Neural network}
\acro{HIO}{Hybrid Input-Output}
\acro{DL}{Deep Learning}
\acro{CNN}{Convolutional Neural Network}
\acro{SSIM}{Structural Similarity Index Measure}
\acro{DIP}{Deep Image Prior}
\acro{XFELs}{X-ray Free-Electron Lasers}
\acro{ER}{Error Reduction}
\acro{GS}{Gerchberg-Saxton}
\acro{MSE}{Mean Squared Error}
\acro{SIREN}{Sinusoidal Representation Networks}
\acro{CDIP}{cross-dimensional implicit priors}
\acro{MAE}{Mean Absolute Error}
\acro{PRTF}{Phase Retrieval Transfer Function}
\acro{FRC}{Fourier Ring Corelation}
\acro{CDIP-S}{CDIP with explicit support} 
\acro{AuNPs}{gold nanoparticles}
\acro{GT}{ground truth}
\acro{MLP}{Multilayer Perceptron}
\acro{Ta}{Tantalum}
\acro{PR}{phase retrieval}
\acro{OTSU}{Otsu's method}
\acro{PDF}{probability density distribution}
\end{acronym}

\setcitestyle{super,comma}

\begin{document}
\title{Constraint-Free Coherent Diffraction Imaging via Physics-Guided Neural Fields}
\author*[1]{\fnm{Zhe} \sur{Hu}}\email{zhe.hu@fysik.lu.se}
\author[1]{\fnm{Zisheng} \sur{Yao}}
\author[1]{\fnm{Yuhe} \sur{Zhang}}
\author[1]{\fnm{Pablo} \sur{Villanueva-Perez}}
\affil[1]{\orgdiv{Synchrotron Radiation Research and NanoLund}, \orgname{Lund University}, \orgaddress{\city{Lund}, \postcode{22100}, \country{Sweden}}}

\abstract{
\ac{CDI} is a lensless imaging technique that enables atomic-resolution imaging of non-crystalline specimens and their dynamics. 
However, its broader implementation has been hindered by the instability and ill-posedness of its reconstruction process, known as phase retrieval, which relies heavily on handcrafted, object-specific constraints.
To overcome the key limitations, we propose \ac{CDIP}, a robust phase-retrieval framework that eliminates the need for such constraints by combining untrained coordinate-based neural fields for static and dynamic reconstructions and a physics-consistent forward model.
We evaluate \ac{CDIP} on simulated and experimental datasets that involve both static samples and dynamic processes, demonstrating that it substantially outperforms classical iterative algorithms and deep-learning baselines in terms of fidelity and stability.
These results highlight a paradigm shift in both static and time-resolved \ac{CDI} reconstruction, providing a broadly applicable framework for coherent imaging modalities such as ptychography and holography, across X-ray, electron, and optical probes.
}
\maketitle

\section{Introduction}
\ac{CDI} is a powerful lensless imaging technique that allows high-resolution structural characterization in materials science, biology, and nanotechnology~\cite{miao2015beyond}. Advances in coherent light sources such as \ac{XFELs} and synchrotrons have extended their reach to ultrafast and nanoscale regimes. At \ac{XFELs}, \ac{CDI} enables femtosecond imaging of single biomolecules before radiation damage occurs~\cite{chapman2006femtosecond,chapman2011femtosecond}. Synchrotron-based \ac{CDI} is widely used to in situ image dynamic processes~\cite{lo2018situ,takayama2021dynamic} and to map strain fields in nanocrystals in diffraction geometry~\cite{miao2015beyond,chen2024correcting}. These experimental advances have intensified demands on phase retrieval algorithms to remain reliable under challenging conditions. In particular, single-shot measurements, fast dynamics, and limited prior knowledge require reconstruction frameworks that are both flexible and robust, without relying on handcrafted constraints or large training datasets.

At the core of \ac{CDI} lies the ill-posed problem of recovering a complex-valued object from intensity-only Fourier-transform measurements, where phase information is lost~\cite{fienup1987phase,miao1999extending,shechtman2015phase}. This results in ambiguities such as global phase shifts, translations, and twin-image artifacts~\cite{marchesini2007invited,zhong2024doubleprior}, and it leaves the solution space severely under-constrained~\cite{Bendory2017FPR}. In dynamic \ac{CDI}, each diffraction pattern may correspond to a distinct state, precluding frame averaging and worsening the reconstruction challenge.
Traditional algorithms such as \ac{ER}~\cite{gerchberg1972practical}, \ac{HIO}~\cite{fienup1987phase}, and compressed sensing~\cite{he2015high} mitigate this by imposing explicit constraints, support, positivity, and sparsity, aiming to narrow the solution space~\cite{marchesini2007invited,elser2003phase}. However, these approaches solve non-convex problems that are prone to local minima and are highly sensitive to initialization and constraint accuracy~\cite{fannjiang2020numerics}. When prior knowledge is scarce or inaccurate, it often fails to recover meaningful solutions.
Deep learning has improved phase retrieval and other inverse tasks through data-driven priors~\cite{sinha2017lensless,rivenson2018phase}, especially in supervised settings~\cite{wang2024usedl,cherukara2018real}. Yet, supervised methods require extensive labeled datasets, paired measurements and reconstructions, which are scarce in real \ac{CDI} due to experimental variability and limited access to ground truth~\cite{barbastathis2019use}.
Unsupervised methods based on the \ac{DIP}~\cite{ulyanov2018deep} have emerged as promising alternatives. \ac{DIP} exploits the natural bias of untrained neural networks toward smooth and coherent structures, serving purely as an architectural prior and offering regularization without the need for external or labeled  data~\cite{wang2024usedl,esfandiari2021deep}. Although effective in Fresnel (near-field) settings~\cite{zhang2021phasegan,bostan2020deep,wang2024usedl}, applying \ac{DIP} to \ac{CDI} remains limited. Existing variants still rely on hand-crafted regularizers or support constraints, restricting their robustness in sparse or complex scenarios~\cite{vu2025pid3net,zhong2024doubleprior,chan2021rapidbcdi,wu2021three}.

Coordinate-based neural representations, commonly referred to as neural fields, have recently gained attention as an effective way to model continuous spatial signals using neural networks that map input coordinates directly to physical quantities such as density, phase, or amplitude~\cite{mildenhall2021nerf,sitzmann2020implicit,zhang20254d}. As an implicit representation of the object, a neural field is not restricted by discrete grids, enabling continuous and resolution-independent reconstruction across the entire domain. These networks are typically implemented as \ac{MLP}, which take spatial (and optionally temporal) coordinates as input and output the corresponding field values. Unlike voxel-based or convolutional architectures that rely on discretized global grids, neural fields provide compact representations with continuous spatial support, allowing efficient memory usage and region-specific inference without processing the full volume.

Here, we propose a novel \ac{CDIP} regularization framework that integrates neural fields with a physics-consistent forward model for robust, constraint-free phase retrieval.
Instead of directly optimizing a 2D complex-valued field, we represent the object as a 3D volumetric neural field parameterized by an untrained neural field, which acts as an implicit prior by restricting the solution to the space of continuous, smoothly varying functions~\cite{rahaman2019spectral}, and is subsequently projected and Fourier-transformed through a differentiable image formation model to match the observed diffraction measurements. 
This 3D representation not only aligns with the underlying physics of image formation but also imposes geometric consistency that helps anchor the solution within a stable reconstruction branch, reducing susceptibility to ambiguities such as twin-image artifacts and spatial translation, which commonly affect 2D optimization.
To further mitigate translation ambiguity, we propose a novel strategy, progressive anchoring, which spatially biases the reconstruction toward the central region of the object plane, thereby eliminating the need for explicit support. Specifically, such an optimization leverages the localized inference property of neural fields: it begins within a small central region and progressively expands outward, guiding the solution toward a consistent spatial alignment.
The framework naturally incorporates time as an additional input coordinate, enabling the encoding of the entire spatiotemporal domain within a unified neural representation. 
This allows the network to model the evolution of continuous objects over time and facilitates temporally coherent reconstructions in dynamic \ac{CDI} scenarios without requiring explicit temporal regularization or handcrafted motion priors.
Together, these contributions represent a paradigm shift in \ac{CDI} reconstruction—from fixed priors and discretized optimization schemes to a compact, flexible, and physically grounded framework driven by coordinate-based neural representations. The proposed method is broadly applicable to both static and dynamic imaging settings and can be naturally extended to other imaging modalities, paving the way for routine, prior-free application of coherent imaging.

\section{Results}
\subsection{Coherent Diffraction Imaging Background}
\ac{CDI} is a lensless imaging technique that reconstructs complex object fields from intensity-only diffraction measurements. 
Let $\mathbf{X} \in \mathbb{C}^{n \times m}$ denote the complex-valued transmission function representing the object’s exit wavefield, where $n$ and $m$ are the number of spatial samples along each dimension. Under the far-field approximation, the observed measurement is the squared magnitude of its Fourier transform~\cite{marchesini2007invited}:
\begin{equation}
\mathbf{Y} = \left| \mathcal{F}(\mathbf{X}) \right|^2 + \boldsymbol{\eta},
\label{eq:forward}
\end{equation}
where $\mathcal{F}$ is the discrete Fourier transform and $\boldsymbol{\eta}$ denotes noise. Since only intensity is recorded, phase retrieval is a nonlinear and ill-posed inverse problem. Classical \ac{CDI} methods mitigate this by enforcing priors, such as known support or non-negativity, and solve a regularized optimization problem\cite{fienup1987phase,gerchberg1972practical}:
\begin{equation}
\min_{\hat{\mathbf{X}}}  \ell\left( \mathbf{Y}, \left| \mathcal{F}(\hat{\mathbf{X}}) \right|^2 \right) + \lambda \cdot \mathcal{R}(\hat{\mathbf{X}}),
\label{eq:pr}
\end{equation}
where $\ell$ is a data fidelity term (e.g., \ac{MAE} or Poissonian log-likelihood), $\mathcal{R}$ encodes priors, and $\lambda$ is a weight that balances the two terms.
Various iterative algorithms have been proposed to solve Eq.~\eqref{eq:pr}, including \ac{ER}~\cite{gerchberg1972practical}, \ac{HIO}~\cite{fienup1987phase}, Relaxed Averaged Alternating Reflections~\cite{luke2004relaxed}, and Oversampling Smoothness ~\cite{rodriguez2013oversampling}. 
These methods alternate between enforcing data consistency in the Fourier domain and applying prior-based constraints in the image domain. 
Their performance relies heavily on accurate prior specification, especially the support mask, and is often sensitive to initialization and noise. 
Despite widespread use, these methods can stagnate or converge to suboptimal local minima, particularly when constraints are misspecified, or measurements are noisy~\cite{marchesini2007invited}. 
These limitations underscore the need for reconstruction methods that are both robust to noise and less reliant on precise constraint specifications. 
In the next section, we present a \ac{CDIP} framework designed to overcome these challenges.

\subsection{CDIP for CDI Reconstruction}
\begin{figure}[htb!]
\centering
\includegraphics[width=0.95\textwidth]{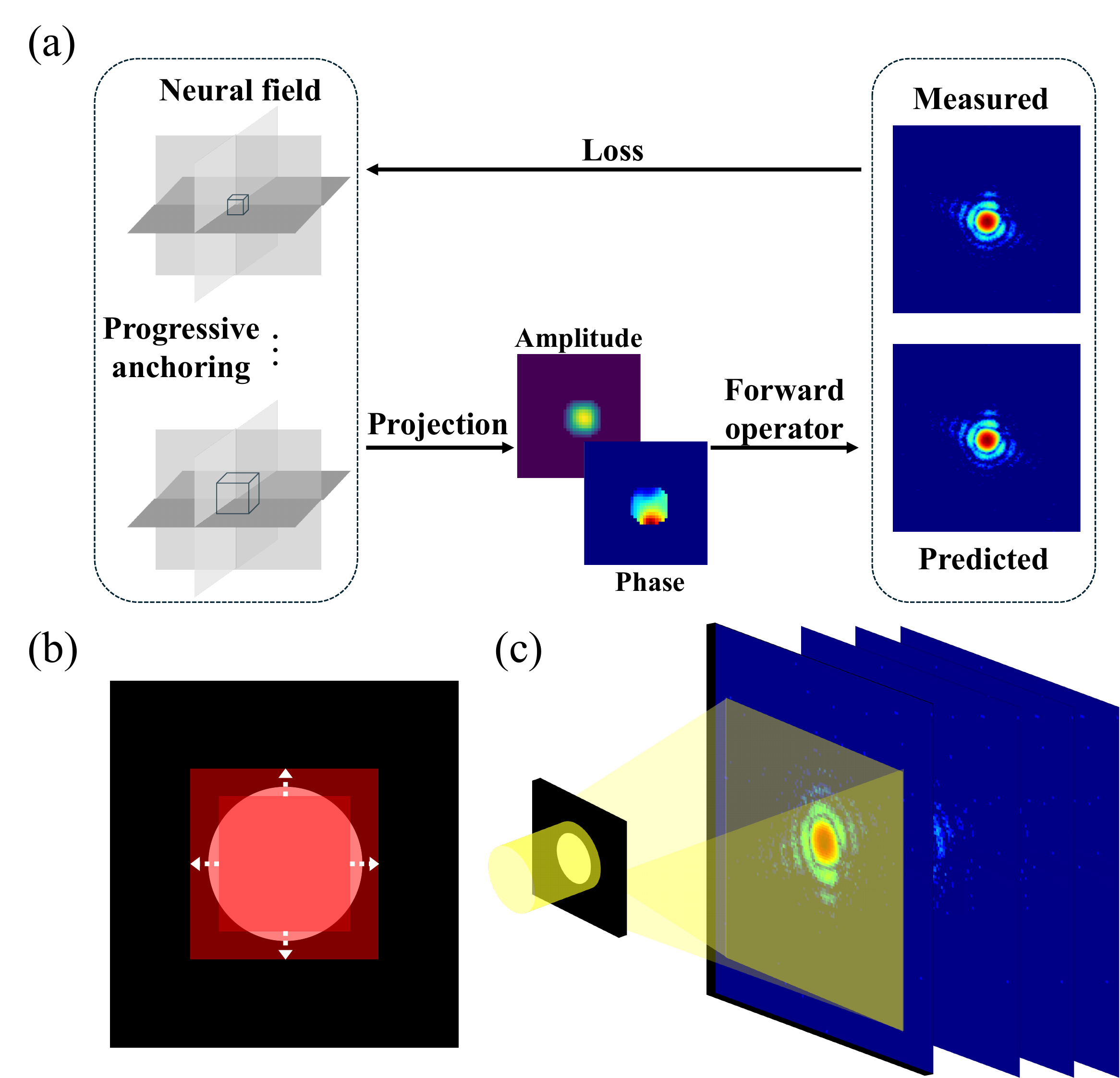}
\caption{Schematic diagram of the \ac{CDIP} concept. (a) Neural fields with a progressive anchoring strategy are used to generate a 3D volume. A differentiable forward operator, comprising a projection and a Fourier transform, is used to simulate predicted diffraction patterns. The neural field is optimized by minimizing the loss between predicted and measured intensities. 
(b) Illustration of the 2D view of the progressive anchoring strategy used in CDIP, where optimization begins from a small central region (red square) projected within the object area.
(c) Illustration of the far-field CDI experiment setup. Coherent X-rays illuminate dynamic objects, and the resulting diffraction patterns are captured sequentially over time. }
\label{fig:CDIP_concept}
\end{figure}

We proposed a neural field representation for \ac{CDI} reconstruction within the \ac{CDIP} framework, leveraging coordinate-based neural priors in combination with a physics-consistent forward model to enable robust, constraint-free phase retrieval, as illustrated in Figure~\ref{fig:CDIP_concept}(a). In this approach, 2D objects are modeled as projections of a continuous 3D neural field, which introduces a strong spatial prior and effectively regularizes the solution space, thereby mitigating the inherent ill-posedness of phase retrieval. 
Crucially, the use of a 3D representation helps to break common symmetries in 2D and 2D+time phase retrieval, such as translational and twin-image ambiguities. By incorporating structural asymmetry and depth context, the 3D neural field anchors the solution within a more stable representation space, reducing the tendency of the optimization to oscillate between multiple equivalent solutions and enabling convergence along a single, consistent solution branch.
Unlike conventional \ac{CNN}-based architectures that operate on discrete global grids, neural fields offer spatial continuity and region-specific inference. This property allows us to focus on optimization in localized areas. We propose a method called progressive anchoring, depicted in Figure~\ref{fig:CDIP_concept}(b), in which the reconstruction begins at the center of the field of view and gradually expands outward. This strategy introduces a natural spatial bias that consistently centers the object, helping to resolve translational ambiguities~\cite{tayal2020inver_prob, Manekar2021breaking} and implicitly suppresses off-object regions without the need for masking. 
Similar to \ac{DIP} methods, neural fields rely on implicit regularization induced by the spectral bias of untrained neural networks, which naturally favor smooth, low-frequency reconstructions~\cite{rahaman2019spectral}. 
To overcome the limited high-frequency expressiveness of standard \acp{MLP} in representing fine-scale structural details, we adopt the \ac{SIREN} architecture~\cite{sitzmann2020implicit} in place of conventional ReLU-based networks. \ac{SIREN} parameterizes spatial fields using sinusoidal activation functions, which are well-suited for modeling high-frequency variations. This choice helps avoid the over-smoothing typically observed in MLP-based reconstructions and enables sharper, more detailed recovery of complex object structures, which is essential for accurate \ac{CDI} phase retrieval.
In addition, we incorporate a perceptual loss to further improve the reconstruction of global morphology. Specifically, this loss is computed from feature activations extracted by a pre-trained VGG network~\cite{johnson2016perceptual}, offering a learned similarity metric that goes beyond pixel-wise intensity differences. Unlike traditional pixel-based losses, the perceptual loss promotes global structural coherence and improves robustness to Poisson noise~\cite{yang2018low,gholizadeh2020deep}, which is particularly common in low-photon-count diffraction measurements~\cite{vu2025pid3net}. 
Notably, the combination of perceptual loss and \ac{SIREN} allows our network to recover both coarse shapes and high-frequency details without relying on tight support masks or hand-crafted priors. When used in conjunction with the progressive anchoring strategy, this design enables fully constraint-free and spatially adaptive phase retrieval.

Furthermore, by treating time as an additional coordinate, the neural field framework naturally extends to dynamic imaging. This temporal extension enables the continuous encoding of information across the entire spatiotemporal space, yielding coherent, frame-consistent reconstructions from diffraction measurements without relying on explicit temporal regularization or sequential data fusion.
Together, these observations highlight the practical advantages of CDIP for solving ill-posed inverse problems in \ac{CDI}, offering a flexible and physically grounded alternative to conventional and data-driven methods.

The object is modeled as a volumetric neural field $\Phi(\mathbf{r})$, where $\mathbf{r} = (x, y, z)$, parameterized by an untrained neural network $\Phi_\theta: \mathbb{R}^3 \rightarrow \mathbb{C}$ that outputs complex-valued amplitude and phase. The forward model comprises a differentiable projection operator $\mathcal{P}$ followed by a Fourier transform $\mathcal{F}$, which simulates the physical image formation process in \ac{CDI}:
\begin{equation}
\hat{\mathbf{Y}} = \left| \mathcal{F} \circ \mathcal{P} (\Phi_\theta) \right|^2.
\label{eq:cross_dim_forward}
\end{equation}
Given the observed diffraction data $\mathbf{Y}$, reconstruction is performed by optimizing network parameters $\theta$ to minimize the discrepancy between measured and predicted intensities:
\begin{equation}
\min_{\theta} \; \ell\left( \mathbf{Y}, \left| \mathcal{F} \circ \mathcal{P} (\Phi_\theta) \right|^2 \right) + \lambda_{\text{perc}} \cdot \ell_{\text{perc}},
\label{eq:cross_dim_loss}
\end{equation}
where $\ell(\cdot, \cdot)$ is a loss in data fidelity (for example, \ac{MAE} or the Poissonian log-likelihood), and $\ell_{\text{perc}}$ is a perceptual loss weighted by $\lambda_{\text{perc}}$. 
Through this formulation, the neural field and physics-consistent forward model jointly constrain the reconstruction to a low-dimensional manifold of plausible 3D structures, providing strong implicit regularization. 

For time-varying reconstructions, as illustrated in Figure~\ref{fig:CDIP_concept}(c), where the object undergoes continuous evolution during image acquisition, we extend the volumetric neural field to a spatiotemporal representation $\Phi_\theta(\mathbf{r}, t)$, where $t$ denotes a continuous temporal coordinate. This formulation enables the network to model smooth temporal dynamics jointly with spatial structure, allowing for coherent reconstruction across time without explicit temporal supervision.
Given a sequence of 2D diffraction measurements $\{\mathbf{Y}_t\}_{t=1}^T$, each corresponding to a different time point, the forward model becomes the following.
\begin{equation}
\hat{\mathbf{Y}}_t = \left| \mathcal{F} \circ \mathcal{P} \left( \Phi_\theta(\mathbf{r}, t) \right) \right|^2+\eta, \quad t = 1, \dots, T.
\end{equation}

To improve temporal stability and spatial smoothness, we introduce a spatiotemporal total variation (TV) regularization term:
\begin{equation}
\ell_{\text{reg}} = \text{TV}_\mathbf{r}\left( \Phi_\theta \right) + \beta \cdot \text{TV}_t\left( \Phi_\theta \right),
\end{equation}
where $\beta$ controls the relative weight of temporal regularization. The final loss aggregates fidelity, perceptual, and regularization terms across time:
\begin{equation}
\min_{\theta} \sum_{t=1}^{T} \ell\left( \mathbf{Y}_t, \left| \mathcal{F} \circ \mathcal{P} \left( \Phi_\theta(\cdot, t) \right) \right|^2 \right) + \lambda_{\text{perc}} \cdot \ell_{\text{perc}} + \lambda_{\text{reg}} \cdot \ell_{\text{reg}}.
\end{equation}
This spatiotemporal formulation enables coherent dynamic reconstruction directly from measurements without requiring frame-to-frame supervision or explicit temporal constraints.
To instantiate the proposed CDIP framework in practice, we adopt the X-Hexplane architecture~\cite{hu2025super}, a structured neural field designed for an efficient and high-resolution representation of 3D and dynamic objects~\cite{chen2022tensorrf,cao2023Hexplane} under X-ray imaging. Implementation details of X-Hexplane can be found in the Methods Section.

\subsection{CDIP on static samples}
\begin{figure}[htb!]
\centering
\includegraphics[width=0.95\textwidth]{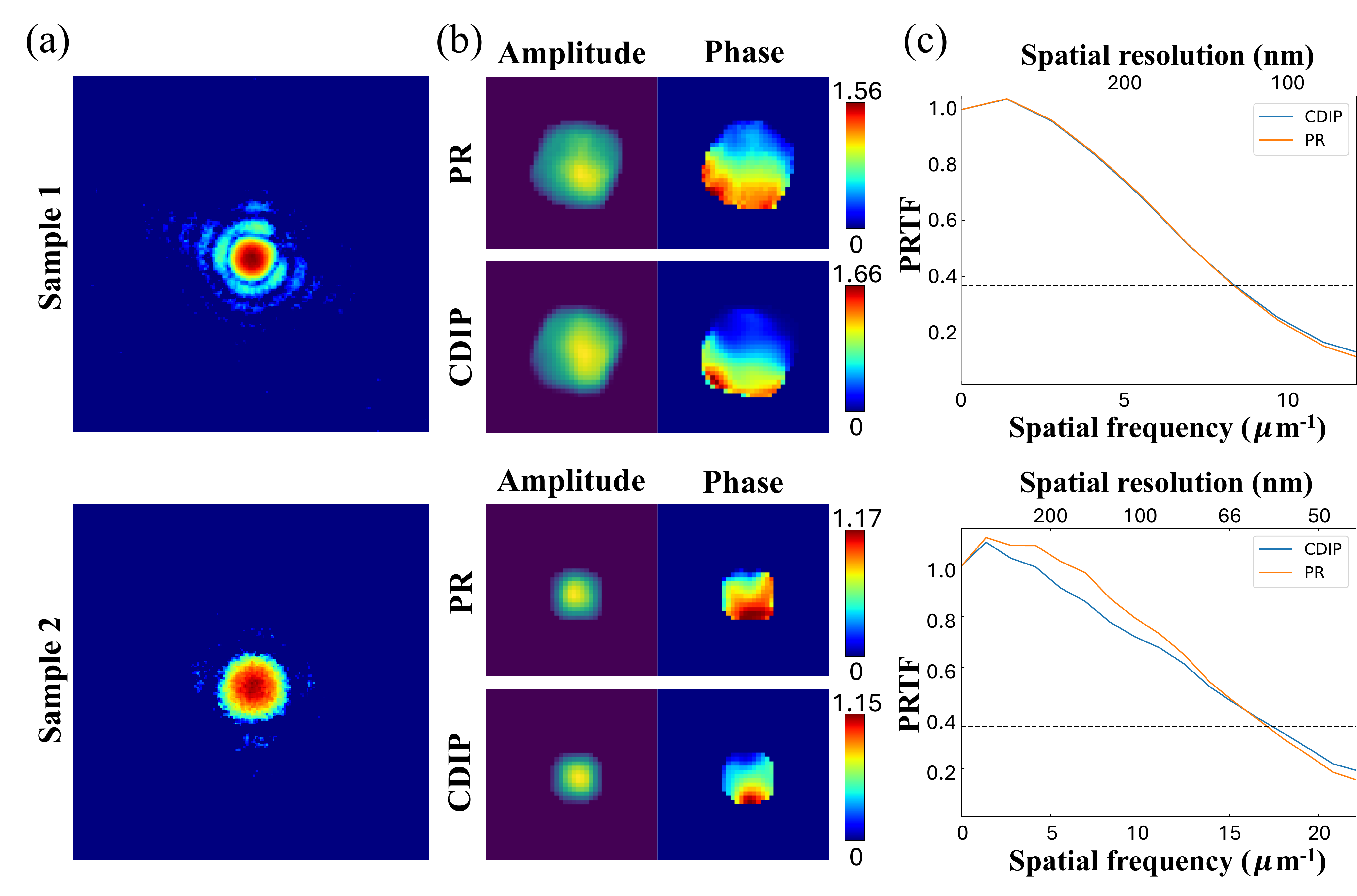}
 \caption{
Comparison of phase retrieval performance between \ac{PR} and the proposed CDIP method. 
(a) Experimentally measured diffraction patterns from two samples. 
(b) Complex-valued fields reconstructed using the \ac{PR} and CDIP methods, with both amplitude and phase components visualized for each sample. The accompanying color bar represents the phase values of each reconstruction. 
(c) \ac{PRTF} curves for the two methods.
}\label{fig:static}
\end{figure}

To validate the effectiveness of our method on experimental data of static objects, we evaluated it using the publicly available diffraction dataset provided in Refs~\cite{bjorling2020three, cxidb:id151}. This dataset consists of coherent diffraction patterns measured from randomly oriented Au nanoparticles of 60 nm diameter using Bragg \ac{CDI}. The X-ray energy was 10~keV, with a sample-to-detector distance of 320~mm and a detector pixel size of 55~$\mu$m. For our static reconstruction experiments, we selected two representative 2D diffraction patterns from the dataset, shown in Figure~\ref{fig:static}(a).  A conventional \ac{PR} algorithm and the proposed CDIP framework are applied to recover the corresponding amplitude and phase distributions. The selected diffraction patterns, with the size of 128x128 pixels, exhibit speckle features and severe Poisson noise, characteristic of real Bragg \ac{CDI} experiments.
This setting provides a real-world test of our framework’s ability to reconstruct high-resolution structural information from experimentally acquired far-field diffraction data. The conventional \ac{PR} algorithm used is a hybrid scheme that combines the updates of HIO~\cite{fienup1987phase} and ER~\cite{gerchberg1972practical} updates, along with Shrinkwrap support refinement. Reconstruction is carried out in the Fourier domain by enforcing the measured amplitude, whereas real-space constraints are applied iteratively in the object domain. Shrinkwrap is performed periodically to adapt the support, and Poissonian log-likelihood loss is monitored to assess convergence. Detailed settings can be found in the Methods Section. In contrast, CDIP is applied without any prior knowledge of the support, symmetry, or training data of the object. A progressive anchoring strategy is used, where the initial optimization region is set to a small 10$\times$10 pixel area and is gradually expanded to around a 30$\times$30 region for both test cases.

Figure~\ref{fig:static}(b) shows the reconstructed complex fields, with individual colorbars indicating the phase range for each method. From the colorbars, it can be observed that the reconstructed phase values of both methods are reasonably well aligned, with only minor discrepancies. While the conventional \ac{PR} approach recovers the overall structure, it suffers from phase discontinuities and boundary artifacts. In contrast, CDIP yields smoother, more coherent reconstructions with better phase continuity, demonstrating its ability to regularize the reconstruction implicitly through its coordinate-based representation.
Figure~\ref{fig:static}(c) shows the quantitative comparison using the \ac{PRTF}, which measures the frequency-dependent agreement between the reconstructed and measured diffraction amplitudes. The threshold $1/e$ (indicated by the dashed line) defines the spatial resolution limit beyond which phase recovery becomes unreliable. The results confirm that both \ac{PR} and CDIP achieve comparable resolution, with estimated resolutions of approximately 125~nm and 60~nm for the two test cases. However, CDIP achieves these results without relying on predefined or adaptively updated tight supports. 

\subsection{Application to dynamic process}
To evaluate the proposed method on dynamic experimental data, we utilized the open-access Tantalum (Ta) test chart dataset, originally published in Ref.~\cite{takazawa2021demonstration,vu2025pid3net,vu2024ta}, as a proof-of-concept for quantitative validation.
In this dataset, the \ac{Ta} test chart is continuously translated while single-shot CDI measurements are recorded, resulting in a video sequence of diffraction patterns. 
Each frame was acquired with an exposure time of 7 ms, allowing time-resolved analysis of the dynamic scattering process. The chart moved at a constant velocity of $340\ \mathrm{nm}/\mathrm{s}$, and five coherent probe modes were provided to account for partial spatial coherence during data acquisition. The predicted diffraction intensity is computed as the incoherent sum of the squared magnitudes of the Fourier transforms of the exit waves from each coherent mode.

We benchmark CDIP against two leading dynamic CDI methods: multi-frame PIE with total-variation regularization (mf-PIE)~\cite{takayama2021dynamic} and the untrained deep-decoding network PID3Net~\cite{vu2025pid3net}.
mf-PIE relies on handcrafted overlapping-frame constraints and explicit TV regularization to enforce temporal smoothness, whereas PID3Net employs a feedforward architecture based on untrained convolutional neural networks with temporal convolution blocks. Both of the methods require a hard support for reconstruction.
In contrast, CDIP, whether in its support-free form or the variant \ac{CDIP-S}, represents the entire space–time volume as a single continuous coordinate-based neural field. This implicit 4D representation naturally enforces projection consistency and spatiotemporal coherence without any explicit temporal regularization, overlapping constraints, or architecture engineering.

\begin{figure}[htb!]
\centering
\includegraphics[width=0.95\textwidth]{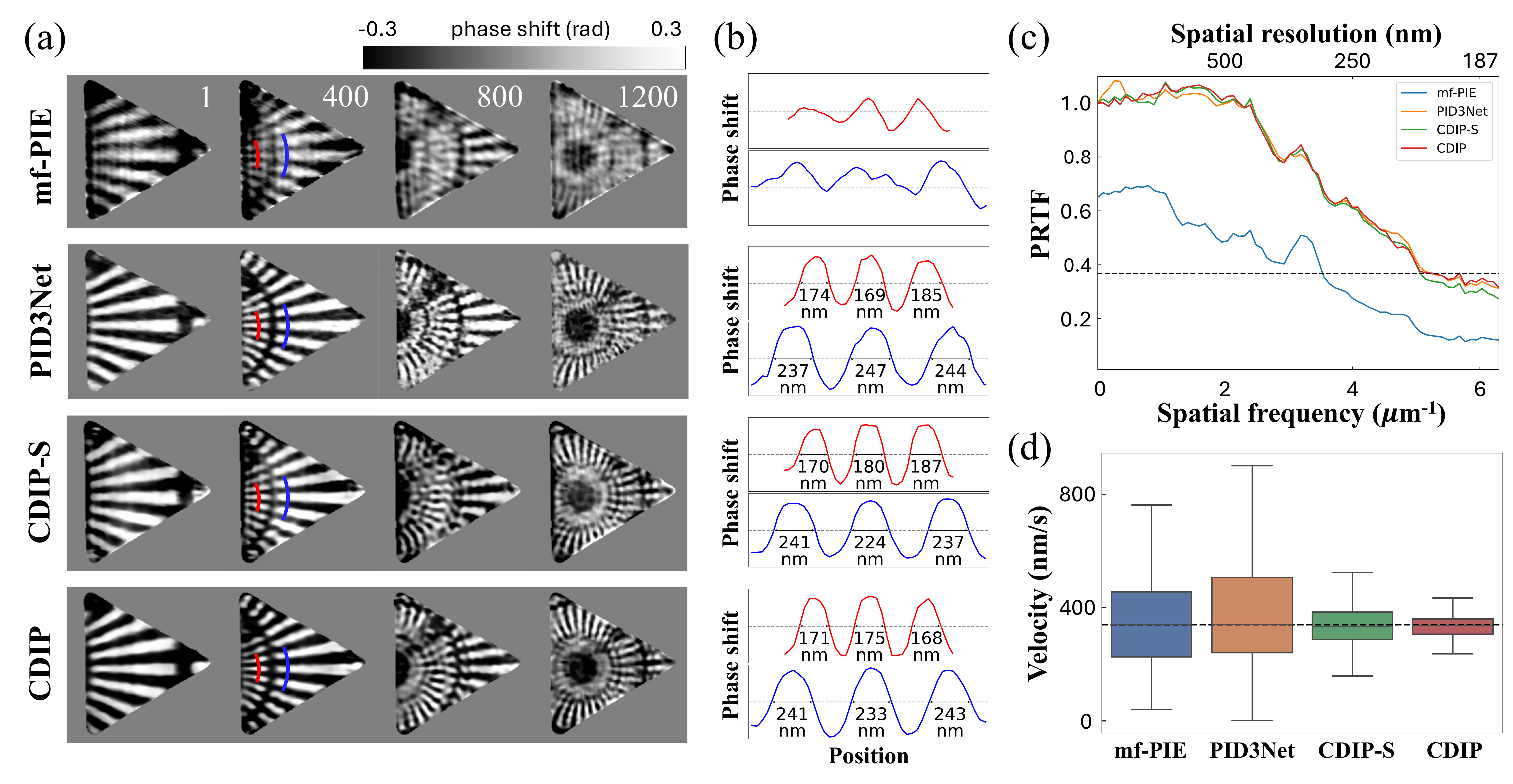}
\caption{
Reconstruction results of the moving \ac{Ta} test chart using different phase retrieval methods. 
(a) Reconstructed phase images at selected time frames (1, 400, 800, 1200) using four methods. 
(b) Phase profiles sampled along two circular arcs in the 400th frame, highlighting the ability of each method to resolve spatially varying patterns with known geometry. 
(c) \ac{PRTF} analysis of the reconstructed phase maps. The dashed horizontal line indicates the resolution threshold at which the PRTF falls below $1/e$, marking the limit of reliable feature recovery. 
(d) Distribution of estimated sample velocities over the first 400 frames derived from each method. The dashed line represents the ground-truth velocity (340 nm/s), and the black lines within each box denote the median of the estimated velocities.
}
\label{fig:Ta}
\end{figure}

Figure~\ref{fig:Ta}(a) presents the phase reconstruction results obtained using four different methods. Visually, the phase images reconstructed by \ac{CDIP-S} and CDIP exhibit clearer patterns and well-defined line absorber shapes in the \ac{Ta} test chart, with stable features over time and significantly reduced background noise. 
Figure~\ref{fig:Ta}(b) shows the retrieved phase values sampled along two circular arcs on the chart. The \ac{Ta} test pattern is arranged in cyclic structures along these arcs, with equal feature widths along each arc and differing widths between the two arcs for both absorbing and transmitting regions. Both CDIP and CDIP-S successfully resolve the spatial variations along these arcs, closely matching the results from PID3Net and accurately capturing the ground-truth geometry of the sample.
Quantitative comparison using the \ac{PRTF} is shown in Figure~\ref{fig:Ta}(c). The mf-PIE method shows the weakest performance, with a spatial resolution limit of around 280 nm, indicating poor recovery of fine features. In contrast, PID3Net, CDIP-S, and CDIP achieve comparable and superior PRTF curves, demonstrating reliable recovery of spatial structures with full-period resolution limits around 200 nm.
Temporal stability is assessed in Figure~\ref{fig:Ta}(d), which shows box plots of the estimated instantaneous velocities of the \ac{Ta} chart over the first 400 frames, computed via auto-correlation between adjacent reconstructed phase images. The dashed line indicates the ground-truth velocity of 340 nm/s. While all methods yield similar median estimates (shown as horizontal black lines), CDIP and CDIP-S exhibit markedly narrower distributions, indicating more stable and temporally consistent reconstructions. In contrast, mf-PIE and PID3Net show significantly larger variance, reflecting frame-to-frame fluctuations. These results highlight the benefit of treating time as an input coordinate within the CDIP framework, which enables coherent dynamic reconstruction without explicit temporal regularization.
Consistent with these findings, Supplemental Movie 1 illustrates that CDIP produces smoother motion and more physically consistent phase evolution over time than the other two approaches. 

Despite the low photon count in the experimental data, our method successfully reconstructs the amplitude of the \ac{Ta} test chart with high clarity, as shown in Supplemental Movie 2 in logarithm scale—a crucial capability for accurately characterizing the sample's intensity distribution. In contrast, this remains a significant challenge for other methods~\cite{vu2025pid3net}.

\begin{figure}[htb!]
\centering
\includegraphics[width=0.95\textwidth,height=0.8\textheight,keepaspectratio]{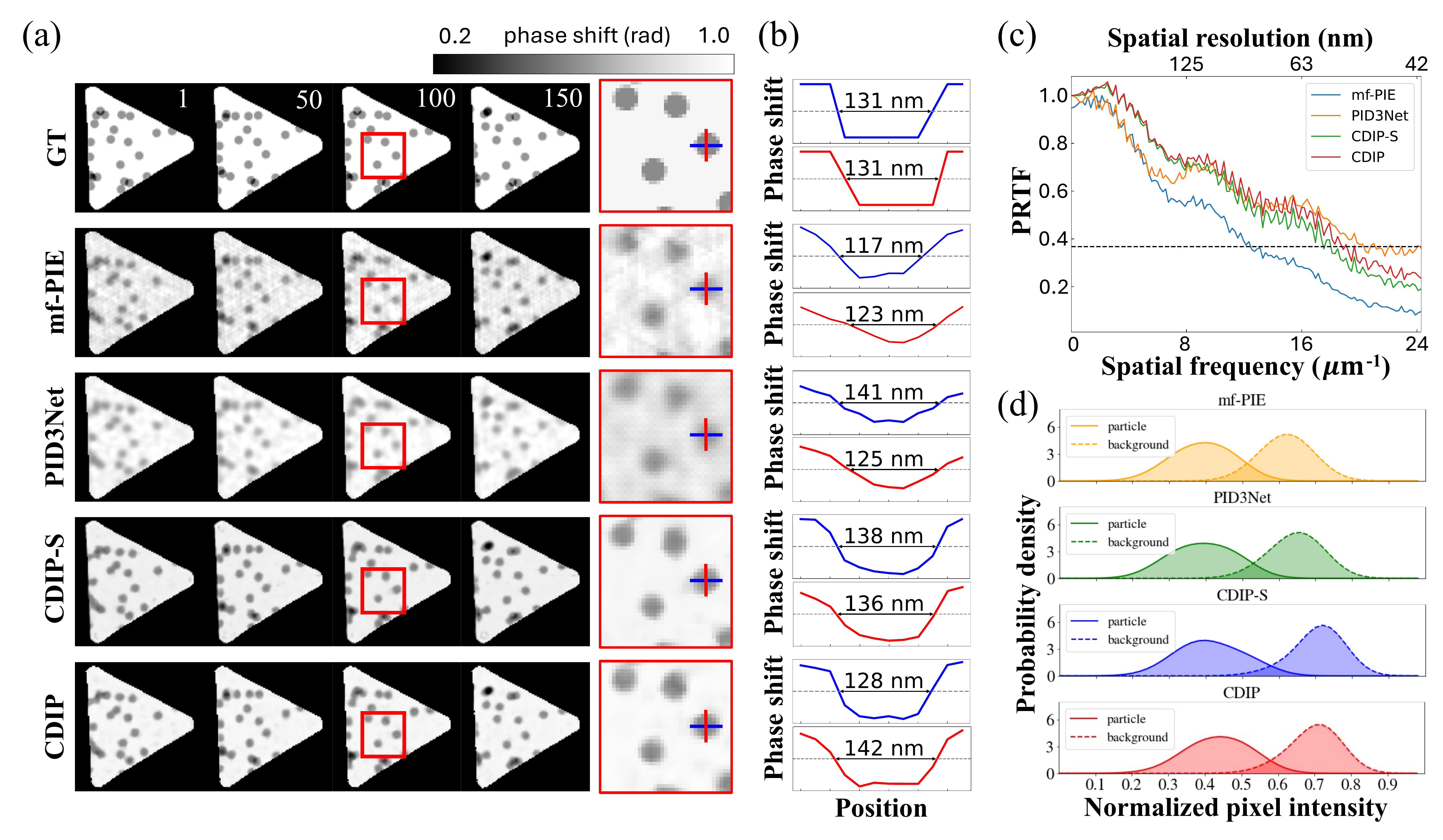}
\caption{
Reconstruction of the dynamic \ac{AuNPs} phase using different methods. 
(a) Phase reconstruction results at selected time frames (1, 50, 100, 150) obtained with different methods. The last column shows zoomed-in views (red boxes) of the highlighted regions at frame 100. 
(b) \ac{PRTF} analysis of the reconstructed phase images. The dashed horizontal line indicates the $1/e$ threshold, marking the cutoff spatial frequency for reliable feature recovery. 
(c) Pixel intensity distributions within the particle region and the solution background in the reconstructed phase images.
(d) Statistical distributions of pixel intensities from the reconstructed phase images, evaluated within the particle and the adjacent solution areas.
}
\label{fig:simu}
\end{figure}

To evaluate the performance of the CDIP in reconstructing dynamic 2D objects, we utilize openly available simulation and experimental datasets provided in Refs.~\cite{vu2025pid3net,vu2024ta}, which model the Brownian motion of \ac{AuNPs} dispersed in aqueous polyvinyl alcohol. These datasets capture dynamic behaviors across a wide spatiotemporal range, from several hundred nanometers to a few micrometers~\cite{takazawa2023coupling}. We begin with numerical simulations, where diffraction patterns are generated by modeling the Brownian motion of \ac{AuNPs} under the assumption of fully coherent plane wave X-ray illumination. To mimic realistic experimental conditions, Poisson noise is added to each diffraction frame, with a total photon count of approximately $3 \times 10^7$ photons per frame.
Figure~\ref{fig:simu}(a) compares dynamic phase reconstruction results obtained using four methods, with the \ac{GT} provided as a reference. All methods successfully capture the coarse motion and global shape evolution of the \ac{AuNPs} ensemble over time. However, as seen in the zoomed-in regions (last column, frame 100), mf-PIE suffers from significant background noise and blurring. At the same time, PID3Net exhibits overly smooth reconstructions, with reduced contrast and boundary sharpness. This degradation stems from the inherent bias of DIP-based architectures toward low-frequency content and the inefficiency of encoding the time axis, limiting their ability to recover fine structural details.
In contrast, both CDIP variants yield high-fidelity reconstructions. \ac{CDIP-S}, aided by tight support constraints, achieves the sharpest boundary delineation. Remarkably, CDIP, despite operating without explicit support, still recovers object structures with high contrast, low background noise, and sharp edges, demonstrating the effectiveness of its implicit spatial regularization and progressive inference scheme. Across all time frames, the CDIP methods consistently preserve object morphology and contrast, validating its robustness under constraint-free and dynamic conditions.
Figure~\ref{fig:simu}(b) compares the line profiles of one representative particle extracted from the zoomed-in regions in Figure~\ref{fig:simu}(a). It is evident that the phase shift ranges for both mf-PIE and PID3Net are smaller than that of the \ac{GT}, and their edge transitions appear smoother. In contrast, CDIP methods accurately recover both the steep edge and the flat regions of the phase profiles, with particularly strong performance observed in the CDIP results. These findings highlight the ability of neural fields to capture high-frequency features and enable robust, high-fidelity phase reconstruction.
Figure~\ref{fig:simu}(c) provides a quantitative comparison via the \ac{PRTF}. CDIP, CDIP-S, and PID3Net achieve the highest spatial resolution, resolving features down to approximately 50~nm (at $1/e$ threshold), significantly outperforming mf-PIE, which degrades around 83~nm. 
To evaluate each method’s ability to distinguish particles from background, an \ac{OTSU} thresholding filter~\cite{otsu1975threshold} is applied to the same region of the reconstructed phase image at the final time point. Pixels with grayscale values below the computed threshold are classified as particles, while the rest are considered background. Figure~\ref{fig:simu}(d) presents the \ac{PDF}) of the grayscale values for particle (solid line) and background (dashed line) regions across different methods. The reconstruction from CDIP variants exhibits a clearer separation between particle and background distributions, indicating enhanced contrast and more accurate particle delineation. The other two methods also demonstrate reasonable contrast, but with less distinct separation.
Supplementary Movie 3 further illustrates the temporal evolution of the reconstructed phase maps. The results of mf-PIE and PID3Net exhibit noisy and unstable backgrounds, which can be attributed to their reliance on information shared between adjacent frames. In contrast, CDIP demonstrates superior frame-to-frame consistency, sharper structural features, and improved noise suppression throughout the sequence.

\begin{figure}[htb!]
\centering
\includegraphics[width=0.95\textwidth,height=0.8\textheight,keepaspectratio]{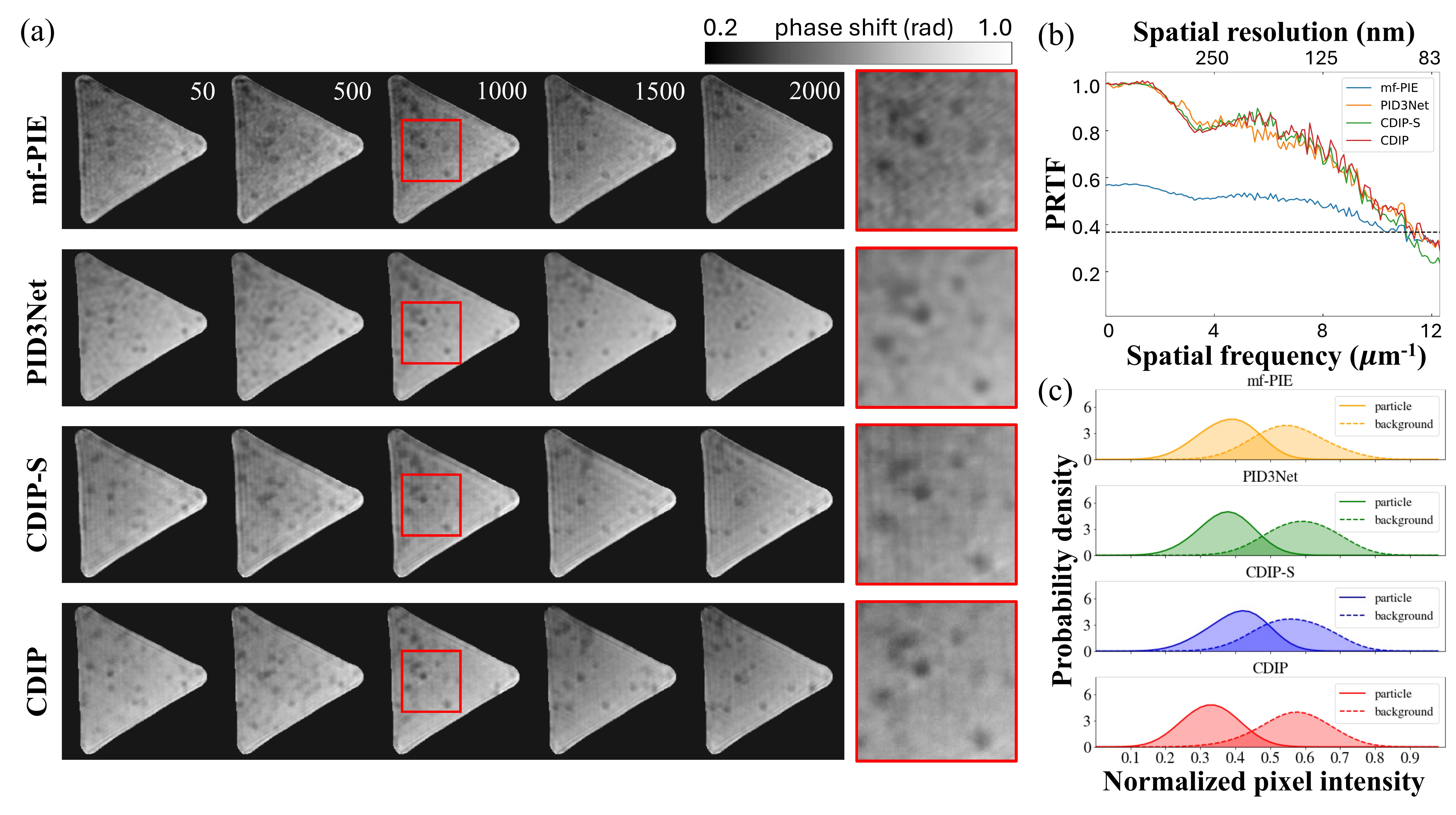}
\caption{
Reconstruction of \ac{AuNPs} phase information from experimentally captured diffraction patterns using different methods. 
(a) Reconstructed phase images at selected time frames (50, 500, 1000, 1500, 2000) using mf-PIE, PID3Net, \ac{CDIP-S}, and CDIP. 
(b) \ac{PRTF} analysis of the reconstructed phase images. The dashed horizontal line indicates the $1/e$ threshold, with the corresponding spatial frequency marked to estimate the resolution limit. 
(c) Distributions of pixel intensities in the reconstructed phase images for the particle and surrounding solution regions.
}
\label{fig:Auexp}
\end{figure}

While previous simulation studies have theoretically demonstrated the effectiveness of the CDIP framework in recovering the dynamic behaviors of \ac{AuNPs}, we further validate its performance under realistic experimental conditions. 
In this third dynamic experiment, we applied CDIP to reconstruct the motion of colloidal \ac{AuNPs} dispersed in a 4.5~wt\% polyvinyl alcohol solution, with particle diameters of approximately 150~nm. The dataset, originally reported in Refs.~\cite{vu2025pid3net, vu2024ta}, includes four coherent probes, which were used to generate the predicted diffraction patterns.
Figure~\ref{fig:Auexp}(a) shows the reconstructed phase images at five representative time points (frames 50, 500, 1000, 1500, and 2000) obtained using four different methods. The last column displays a zoomed-in view of the region highlighted by the red box in frame 1000. The results of mf-PIE show pronounced blurring and elevated background noise, which limit its ability to resolve particle boundaries or track motion reliably. PID3Net offers improved noise suppression but still suffers from reduced contrast and overly smooth boundaries.
In contrast, both CDIP variants produce sharper and cleaner reconstructions. CDIP, in particular, captures particles with high fidelity and temporal consistency across frames, despite the absence of explicit support constraints. CDIP-S also improves on baselines but introduces edge artifacts and local distortions. This degradation is likely due to the fixed hard support and Poisson noise assumptions enforced during optimization, which may conflict with the true noise characteristics of the experimental data and reduce robustness when prior knowledge is inaccurate.
The PRTF analysis in Figure~\ref{fig:Auexp}(b) quantitatively supports these findings: CDIP and CDIP-S exhibit resolution recovery comparable to baseline methods. 
\ac{OTSU} method was also applied to experimental results, and Figure~\ref{fig:Auexp}(c) shows the \ac{PDF} of the particle and background regions. CDIP and PID3Net achieve the clearest separation between distributions, indicating robust phase reconstruction under noisy, low-photon count conditions. CDIP-S performs worse on this metric, consistent with the structural artifacts observed earlier.
Supplemental Movie 4 further visualizes the reconstructed dynamics, confirming that CDIP provides smoother temporal evolution, stronger phase contrast compared to all baselines.
Together, the results confirm the practical effectiveness of CDIP for dynamic CDI under experimental noise and demonstrate the robustness of CDIP in unconstrained reconstruction scenarios.

\section{Discussion}
In this work, we introduce \ac{CDIP}, a phase-retrieval framework for constraint-free \ac{CDI} reconstruction from intensity-only diffraction data. \ac{CDIP} employs untrained coordinate-based neural fields to represent static and dynamic complex-valued structures, which are jointly optimized through a physics-consistent forward model, thereby unifying structural continuity, compactness, and physical interpretability within a single differentiable framework. A key advantage of \ac{CDIP} is its implicit neural prior, which naturally biases solutions toward smooth, spatially coherent structures while enabling local inference—achieving high-fidelity reconstruction without supervision, initialization, or handcrafted constraints.

Quantitative benchmarks demonstrate that \ac{CDIP} achieves significantly superior reconstruction fidelity and stability relative to classical phase retrieval, state-of-the-art neural network baselines. Notably, \ac{CDIP} enables temporally coherent reconstructions from time-resolved diffraction data without requiring frame-to-frame coupling or explicit temporal regularization—a capability that prior methods can only approximate through strong constraints. It also excels in recovering high-frequency details, suppressing background artifacts, avoiding interference between time frames, and producing stable dynamic reconstructions across varying photon counts and sampling regimes. Quantitative metrics, including PRTF, \ac{OTSU}, and velocity estimation, support its effectiveness in both spatial accuracy and temporal consistency.
Despite its strengths, the current implementation based on the X-Hexplane architecture exhibits limitations under certain dynamic conditions. For example, in the Ta chart experiment, the reconstruction stability degrades between frames 800 and 1000, as demonstrated in Supplemental Movie 5, due to the exit of high-frequency features and the entrance of new low-frequency content, which disrupts the feature-sharing mechanism of X-Hexplane. Similarly, in the experimental AuNPs dataset, large inter-frame displacements can lead to slight particle deformation, as the network attempts to interpolate motion smoothly. These observations suggest that while X-Hexplane serves as an efficient backbone for dynamic \ac{CDI}, future work may explore alternative architectures that preserve temporal continuity while improving robustness to abrupt object motion under the guidance of CDIP.

In summary, CDIP provides a robust and constraint-free solution for high-fidelity phase retrieval in both static and dynamic CDI. 
By removing the need for handcrafted constraints, initialization, and massive training data, it addresses a key algorithmic bottleneck that has limited the broader adoption of CDI at synchrotrons and X-ray free-electron lasers.
More broadly, CDIP exemplifies a general principle: embedding exact physical models within coordinate-based neural representations to solve ill-posed inverse problems. 
This paradigm extends naturally to ptychography~\cite{li2024x,miao2025computational}, Bragg CDI~\cite{yang2013coherent,yao2022autophasenn}, 3D/3D+time coherent imaging~\cite{clark2012high,scheinker2020adaptive}, and other coherent imaging modalities across the X-ray~\cite{weber2022evolution}, electron~\cite{jiang2018electron,chen2021electron}, and optical domains~\cite{xu2024high}—paving the way for a new generation of universal, constraint-free computational imaging at the nanoscale.

\section*{Methods}
\subsection*{X-Hexplane for CDI}
Here, we adopt X-Hexplane as the untrained neural architecture for phase retrieval from diffraction patterns. X-Hexplane is a coordinate-based neural representation designed to model a refractive index field $n(\mathbf{x}, t)$ over spatial coordinates $\mathbf{x} = (x, y, z)$ and time $t$ by factorizing the underlying 4D function into a set of learnable low-rank tensor planes~\cite{chen2022tensorrf,cao2023Hexplane,hu2025super}. This structure enables efficient and scalable querying of volumetric and temporal information, making it particularly suitable for large-scale inverse problems. Although originally developed for near-field imaging, we adapt X-Hexplane to the far-field \ac{CDI} setting by integrating a physics-consistent forward model based on the Fourier transform.
In the static setting (i.e., single-shot \ac{CDI}), we configure X-Hexplane to represent a time-independent 3D field $\Phi_\theta(\mathbf{r})$ and reconstruct it by optimizing against a single measured diffraction pattern. 
For dynamic reconstructions, X-Hexplane is extended to handle time-varying objects by encoding time $t$ as an additional input coordinate. Instead of training separate models for each frame, a single set of tensor features is shared across time steps. This design allows the model to learn temporally coherent object dynamics while avoiding redundancy and frame-wise overfitting. By jointly encoding space and time within a unified spatiotemporal coordinate system, X-Hexplane enables smooth and consistent recovery of evolving structures, a crucial property for dynamic \ac{CDI}.
We incorporate progressive anchoring into X-Hexplane by selecting a reference area. This encourages the network to localize and refine the object center early in optimization, functioning as an implicit support prior to the need for hand-made masks. An optional outer boundary, typically set to half the size in each dimension, can be applied to limit computational overhead by excluding unnecessary zero-value voxels from training.
X-Hexplane supports multiple activation and loss functions to accommodate varying imaging scenarios. For activation function, both Leaky ReLU~\cite{xu2020reluplex} and \ac{SIREN} are available, while the loss functions include Poissonian log-likelihood and \ac{MAE}, allowing adaptation to different noise models and reconstruction objectives.
Our implementation is based on PyTorch 1.6.0 with Python 3.8.8. Training and inference were performed on an NVIDIA A100 GPU with 80 GB of memory. For 128$\times$128 single-shot images, reconstruction typically required around 1 min for 2000 epochs. For dynamic 2D cases with an image size of 384$\times$384 pixels in 2000 time steps, the training required approximately 2 hours. Inference (2D rendering) took around 0.2 seconds per frame.

\subsection*{Implementation details of the standard phase retrieval}
As a baseline comparison, we implemented a traditional hybrid \ac{PR} algorithm combining the HIO, ER, and Shrinkwrap strategies~\cite{latychevskaia2018iterative}. The algorithm is initialized with a random complex-valued guess and iteratively updates the reconstruction using the measured Fourier amplitude and an initial real-space support constraint. Every ten iterations, ER updates are applied to promote conservative convergence, while HIO is used otherwise with a relaxation factor $\beta = 0.9$. The final 20\% of iterations exclusively apply ER to ensure stability.
Poisson noise statistics are considered in the loss calculation. Shrinkwrap is activated every 100 iterations by applying a Gaussian filter ($\sigma = 2.0$) to the current amplitude estimate, followed by thresholding and morphological opening to adaptively update the support. The reconstruction enforces non-negativity on the real part of the object in each iteration.
\section*{Data Availability}
The static data is available at Ref~\cite{cxidb:id151}. The dynamic datasets of the moving Ta test chart and gold nanoparticles are available at Ref~\cite{vu2024ta}.
\section*{Code Availability}
Code has been made publicly available at: available upon publication.

\section*{Acknowledgments}
Z.H thanks Z. Matej for his support and access to the GPU-computing cluster at MAX IV. 
This work has received funding from ERC-2020-STG 3DX-FLASH 948426 and the HorizonEIC-2021-PathfinderOpen-01-01, MHz-TOMOSCOPY 101046448.

\section*{Author Contributions}
Z.H. and P.V.-P. conceived and conceptualized CDIP. Z.H., Z. Y., Y. Z., and P.V-P. developed and contributed to the neural network framework and physical formulation of the problem. 
Z.H. performed the data analysis. 
P.V.-P. supervised the research.
Z. H. and P. V.-P. wrote the article with input from all the coauthors.

\section*{Competing interests}
The authors declare no competing interests.
\bibliography{reference}
\end{document}